%% file: hfBalancingModelPaper.tex
\documentclass[a4paper,11pt]{article}
\usepackage{jheppub} % for details on the use of the package, please see the JINST-author-manual
\usepackage{lineno}
\usepackage{booktabs}
\usepackage{tabularx}
\usepackage{array}
\usepackage{placeins}
\usepackage{graphicx}
\usepackage{lineno}
\usepackage{xspace}

\usepackage{color}
\definecolor{orange}{rgb}{1,0.5,0}
\definecolor{darkgreen}{rgb}{0.0, 0.6, 0.22}

\usepackage{hyperref}
\usepackage{amsmath}

\title{Balancing heavy flavour: Quantum number compensation in the hadronisation of charm and beauty particles in proton–proton collisions
}

\author[a]{I.~Pardo Zambrana,}
\emailAdd{i.pardo@student.maastrichtuniversity.nl}
\author[b]{P.~Veen,}
\emailAdd{Paul.Veen@cern.ch}
\author[a,c]{K.K.~Vos,}
\emailAdd{k.vos@maastrichtuniversity.nl}
\author[a,c]{P.~Christakoglou.}
\emailAdd{Panos.Christakoglou@maastrichtuniversity.nl}

\affiliation[a]{Universiteit Maastricht,
Maastricht, The Netherlands}

\affiliation[b]{Paris-Saclay IRFU CEA,
Gif-Sur-Yvette, France}

\affiliation[c]{Nikhef,
Amsterdam, The Netherlands}

\abstract{The hadronisation of heavy quarks remains one of the least understood processes of Quantum Chromodynamics (QCD). While the production of heavy quark pairs can be calculated perturbatively in QCD, the subsequent formation of hadrons involves non-perturbative processes. The description of these processes is predominantly model dependent and sensitive to the underlying mechanisms of colour reconnection and string fragmentation. In particular, the conservation of flavour, baryon number and charge imposes constraints on how heavy-flavour hadrons are formed and balanced within the final state.

In this work, we study how the quantum numbers associated with charm and beauty production are balanced in proton-proton collisions, using events simulated with the PYTHIA-8 event generator.  Starting from perturbatively produced ($c\overline{c}$) and ($b\overline{b}$) pairs, we examine how the formation of heavy-flavour mesons and baryons, including $D^0$, $B^+$, $\Lambda_c^+$, and $\Lambda_b^0$, is accompanied by compensating particles. The analysis focuses on correlations between the identified heavy hadron and the hadronic system responsible for balancing its conserved quantum numbers. We compare three hadronisation configurations that are implemented in PYTHIA: the Monash tune, colour-reconnection models incorporating string junctions (SJ), and close-packing (CP) approaches. We find that the SJ and CP tunes show a substantially enhanced contribution of baryons to the charm and beauty balancing yields compared with the Monash tune; this enhancement becomes more pronounced in high-multiplicity events.

This study introduces a framework for characterising quantum-number compensation in heavy-flavour production and provides a new way to test hadronisation dynamics using identified charm- and beauty-hadron correlations, which can be explored at current and future collider experiments.}

\begin{document}
\maketitle
\raggedbottom

%________________________________________________%
\input{sections/Introduction}
\input{sections/Observables}
\input{sections/Model}

\input{sections/Results}

\input{sections/Summary}
%------------------------------------------------%

%------------------------------------------------%
\section*{Acknowledgements}
Support from the following Dutch Research Council (NWO) research grants is gratefully acknowledged by P.C.: NWO-XL grant OCENW.XL21.XL21.038 and NWO-M1 grant OCENW.M.24.214. The work of K.K.V. is partly supported by the Dutch Research Council (NWO) as part of the project Solving Beautiful Puzzles (VI.Vidi.223.083) of the research programme Vidi.

We thank Sophia Meewis and McKenzie Pedro for their contributions during the early stages of the project. We also thank Rik Spijkers for his essential contributions to early developments of the analysis and providing supervision whenever necessary.
%------------------------------------------------%

\newpage
\bibliographystyle{JHEP}
\bibliography{hfBalancingModelPaperRefs}{}

\clearpage
\appendix
\input{sections/Appendices}

\end{document}

%% file: sections/Introduction.tex
%________________________________________________%
\section{Introduction}
\label{Sec:Introduction}

The transition from coloured quarks and gluons to colour-neutral, colour-singlet hadronic states, commonly referred to as hadronisation, remains one of the central open questions in Quantum Chromodynamics (QCD)~\cite{Andersson:1983ia,Webber:1994zd,Collins:1989gx,Metz:2016swz}. While the production of energetic partons can be described using perturbative techniques, the subsequent confinement of these partons into observable hadrons occurs at energy scales where the strong coupling becomes large and perturbative calculations are no longer applicable \cite{Gross:2005kv,Politzer:1973fx,Ioffe:2010zz,Greensite:2011zz}. As a result, hadronisation is usually described through phenomenological models that incorporate the essential principles of QCD while relying on experimentally constrained parameters.

The most widely used framework for describing hadronisation in high-energy collisions is the Lund string model~\cite{Andersson:1997xwk,Andersson:2002ap}. In this picture, the colour field between separating colour charges is represented by a string with approximately constant tension. As partons move apart, potential energy is stored in the string until it becomes energetically favourable to produce new quark--antiquark pairs from the vacuum. This consequently fragments the string into smaller pieces that subsequently form hadrons. Implemented in the event generator PYTHIA~\cite{Sjostrand:2014zea,Bierlich:2022pfr}, the Lund string model has been remarkably successful in describing a wide range of observables in lepton and hadron collisions.

The environment of proton--proton collisions, however, introduces additional complexity compared to the simple fragmentation of a single colour-singlet system. Multiple parton interactions (MPI), beam remnants, and overlapping colour fields generate a dense environment of colour fields. To account for these effects, PYTHIA incorporates colour reconnection (CR) mechanisms that allow partons originating from different partonic scatterings to reorganise their colour topology before hadronisation. In the Monash tune, reconnections are performed so as to minimise the overall string length, leading to an improved description of global event properties~\cite{Bierlich:2015rha,Skands:2014pea}. Despite its success, the default string fragmentation framework, even with the inclusion of colour reconnection schemes, encounters significant challenges when confronted with measurements of production of different particle species at the Large Hadron Collider (LHC). In particular, the observed enhancement of strange-hadron and baryon production with increasing event activity cannot be fully reproduced by conventional string fragmentation and colour reconnection models~\cite{ALICE:2016fzo,Altmann:2025afh,CMS:2016zzh,ATLAS:2024nbm}. This has motivated the development of extensions that incorporate additional non-perturbative dynamics.

One such extension is the Beyond-Leading-Colour colour reconnection model~\cite{Christiansen:2015yqa}, which introduces string junction (SJ) topologies. Junctions correspond to Y-shaped colour structures in which three string pieces meet at a common point, providing a natural mechanism for baryon production beyond the traditional diquark picture. The presence of junctions modifies not only baryon yields but also the correlations and balancing of conserved quantum numbers throughout the event.  
More recently, close-packing (CP) approaches have been developed to provide a geometrically inspired description of overlapping colour fields and their collective hadronisation~\cite{Altmann:2025afh}.

The importance of these mechanisms has been highlighted by a series of measurements performed at the LHC. In particular, the observation by the ALICE Collaboration of a strong multiplicity-dependent enhancement of strange and multi-strange hadron production in proton--proton (pp) collisions~\cite{ALICE:2016fzo} demonstrated that phenomena traditionally associated with larger collision systems are already present in small systems. Similar trends have been observed for baryon-to-meson ratios \cite{ALICE:2021npz,ALICE:2020wla,ALICE:2023wbx,ALICE:2020wfu,ALICE:2017thy,ALICE:2021rzj,ALICE:2021psx,ALICE:2021bli,ALICE:2017dja,ALICE:2022cop,LHCb:2023wbo}, indicating that hadron production depends not only on the flavour content of the produced quarks but also on the details of the hadronisation mechanism. Furthermore, ALICE measurements of inclusive $\mathrm{J}/\psi$ production as a function of charged-particle multiplicity revealed a significantly stronger-than-linear increase of the charmonium yield with event activity~\cite{ALICE:2020msa}. This observation points to an important role of multiple parton interactions and suggests a non-trivial connection between heavy-flavour production and the global properties of the event. Together, these measurements have established identified-particle production, spanning both light- and heavy-flavour sectors, as a sensitive probe of non-perturbative QCD dynamics and the mechanisms governing hadron formation. 

Most studies up to now have focused on light-flavour hadrons, where both the production mechanism and the hadronisation process contribute to the observed final-state particle yields. This complicates the interpretation of baryon-to-meson ratios and quantum-number correlations, since the underlying partonic production rates are themselves difficult to constrain experimentally. Heavy flavour, however, offers a unique opportunity in this respect. Charm and beauty quarks are predominantly produced in hard partonic scatterings at energy scales where perturbative QCD is applicable, and their production cross sections can, in principle, be calculated from first principles. Consequently, heavy-flavour hadrons provide a cleaner laboratory for isolating hadronisation effects from dynamics associated with partonic production. 

In this work, we investigate how conserved quantum numbers are balanced during the hadronisation of heavy quarks in proton--proton collisions at LHC energies. Using PYTHIA-8, we compare the expectations of the Monash tune, junction-based colour reconnection, and close-packing hadronisation models. Starting from produced $c\overline{c}$ and $b\overline{b}$ pairs, we study how the formation of heavy-flavour mesons and baryons is accompanied by balancing particles elsewhere in the event using a two-particle correlation technique described in Section~\ref{Sec:Observables}. The tunes of PYTHIA-8 and the relevant parameters used for the production of the data samples are described in Section~\ref{Sec:Model-and-Analysis}. The results, presented in Section~\ref{Sec:Results}, examine the heavy-flavour correlations associated with identified charm- and beauty-hadron production. These correlations provide new insight into the microscopic mechanisms governing hadronisation and indicate that balancing observables can discriminate between different models of non-perturbative QCD. We conclude in Section~\ref{Sec:Summary}. Additional details are presented in the appendices. 

%% file: sections/Observables.tex
\section{Two-particle correlations and quantum-number balancing}
\label{Sec:Observables}

Traditionally, hadronisation models are constrained through measurements of identified-particle yields and yield ratios. Examples include baryon-to-meson ratios such as $\Lambda/\mathrm{K}^0_{\mathrm{S}}$, $\Lambda_{c}/D^0$, or $\Lambda_{b}/B^+$, and measurements of strangeness enhancement as a function of event multiplicity. Such observables have proven highly valuable for identifying deficiencies in hadronisation models and motivating the development of new mechanisms such as colour reconnection, junction formation, and rope or close-packing effects. However, yield measurements provide only integrated information on the final state and are therefore often not suitable to identify the microscopic origin of the observed hadronisation patterns. Different hadronisation mechanisms may lead to similar integrated yields while producing very different correlation patterns among the particles generated in the same event.

Two-particle correlation techniques provide a more differential probe of hadronisation dynamics. Rather than measuring only the yield of a given hadron species, one studies the yield of associated particles in the presence of a selected trigger species. The distribution of associated particles can be investigated as a function of kinematic variables such as the relative azimuthal angle $\Delta\varphi$ or the pseudorapidity difference $\Delta\eta$. Since conservation laws operate on an event-by-event basis, the balancing partners of a given particle contain direct information about the mechanisms responsible for its production.

A particularly powerful application of this technique is the study of conserved quantum numbers. Electric charge, baryon number, strangeness, charm and beauty must be conserved globally in every collision event. Consequently, the production of a particle carrying a non-zero quantum number requires the existence of one or more balancing particles elsewhere in the event. The simplest example is electric charge conservation. If a positively charged trigger particle is selected, one may construct correlation functions with positively and negatively charged associated particles. The unlike-sign correlation function contains contributions from both charge-conserving processes and background sources unrelated to charge conservation. To isolate the balancing contribution, it is common to construct a charge-balance observable of the form \cite{Bass:2000az}

\begin{equation}
B_{\rm charge}(\Delta\eta,\Delta\varphi) = 
C_{+-}(\Delta\eta,\Delta\varphi)-
C_{++}(\Delta\eta,\Delta\varphi),
\label{eq:charge-balance}
\end{equation}

\noindent where $C_{+-}$ and $C_{++}$ denote the unlike-sign and like-sign correlation functions, respectively, defined as

\begin{align}
C_{+\mp}(\Delta\eta,\Delta\varphi) = \frac{1}{N_{\text{triggers}}}\frac{d^{2}N_{+\mp}}{d(\Delta\varphi) d(\Delta\eta)} \ , 
\label{eq:correlation-function}
\end{align}
where $N_{\text triggers}$ is the number of trigger particles. 
\noindent Equivalent definitions can be constructed using negatively charged trigger particles or by averaging positive and negative trigger samples.

The subtraction of the $C_{++}$ function removes a large fraction of the combinatorial background and highlights the particles responsible for balancing the trigger charge. The amplitude, width and shape of the resulting distribution provide information on the dynamics of local charge conservation during hadronisation. The same concept can be extended to other conserved quantum numbers. For strangeness, one may investigate correlations between strange and anti-strange hadrons. For baryon number, correlations between baryons and anti-baryons reveal how baryon-number conservation is implemented during fragmentation. Such measurements have already been proposed as sensitive probes of junction topologies~\cite{Lonnblad:2023kft}, since baryons produced through junction formation are expected to exhibit balancing patterns that differ substantially from those generated through the traditional diquark mechanism (see also \cite{ALICE:2025ytq}).

Heavy flavour, on the other hand, offers an advantageous feature for these studies since charm and beauty quarks are produced predominantly in hard partonic scatterings as $c\overline{c}$ and $b\overline{b}$ pairs. The production process therefore establishes a well-defined heavy-flavour quantum number at the partonic level before hadronisation takes place. During fragmentation, these heavy quarks are confined into charm- and beauty-hadron species whose quantum numbers must ultimately be balanced elsewhere in the event. For example, the production of a $D^0$ meson containing a charm quark requires the corresponding anti-charm quantum number to be carried by another hadron. Similarly, the formation of a $\Lambda_{c}^+$ baryon involves not only charm conservation but also the redistribution of baryon number among the remaining hadrons produced during fragmentation.

The nature of these balancing partners is expected to depend strongly on the hadronisation model. In conventional string fragmentation, quantum-number compensation tends to occur locally along the fragmenting string. Junction topologies may redistribute baryon number over larger regions of phase space and alter the flavour composition of balancing particles. Likewise, close-packing mechanisms may modify the probabilities for producing light quarks and diquarks, potentially changing both the species and kinematic distributions of the balancing hadrons. Consequently, two-particle correlation measurements provide information that is largely inaccessible through inclusive yield measurements alone. Mixed-species charge and baryon balance functions have been studied in PYTHIA~\cite{Pruneau:2024jpa}, as have charm-hadron correlations and balance functions~\cite{Sheibani:2025iuh}.

In the present work, we apply quantum-number balancing to charm and beauty hadrons and compare the predictions of the Monash, SJ and CP configurations. By selecting identified charm and beauty mesons and baryons as trigger particles and studying their correlations with associated heavy-flavour particles, we investigate how charm and beauty are compensated in different PYTHIA hadronisation models. The resulting correlation patterns provide a direct probe of the microscopic mechanisms governing heavy-flavour hadronisation.

%% file: sections/Model.tex
\FloatBarrier
\section{Simulation model and analysis}
\label{Sec:Model-and-Analysis}

Three PYTHIA-8 configurations, referred to as tunes later in the text, are considered: Monash, String Junctions (SJ) and Close-Packing (CP). These configurations differ in their parameter settings and hadronisation prescriptions, allowing us to investigate the sensitivity of the measured observables to different modelling approaches.

The simulations were performed with PYTHIA~8.317 for pp collisions at a centre-of-mass energy of $\sqrt{s} = 13.6$~TeV, corresponding to the collision energy of LHC Run~3. For each tune, a sample of 100 million successfully generated events was produced, resulting in a total of 300 million events across all three tunes. The samples are enriched in heavy flavour and are generated using PYTHIA's hard-process selections for charm and beauty production. Specifically, both \texttt{HardQCD:hardccbar} and \texttt{HardQCD:hardbbbar} are enabled, with \texttt{PhaseSpace:pTHatMin = 2}, requiring the transverse momentum of the hard scattering to satisfy ($\hat{p}_{\mathrm{T}}\geq 2$~GeV/$c$). This selection enhances the heavy-flavour content of the simulated samples and provides sufficient statistical precision for the study of identified heavy-hadron correlations. Parton showers, multiple parton interactions (MPI) and beam remnants are retained in the event generation. Nevertheless, the selected samples do not represent the inclusive mixture of heavy-flavour production mechanisms in pp collisions. The same hard-process selection is applied to all three tunes, providing a common basis for their comparison. Consequently, the results presented in this work are conditional on the specified heavy-flavour production selection.

To focus on heavy hadrons at the production stage, the decays of all charm- and beauty-containing hadrons are disabled, including those of excited and hidden-heavy-flavour states. This treatment also excludes contributions from strong and electromagnetic feed-down to the corresponding ground-state hadrons. Excited heavy-flavour hadrons are therefore retained as distinct species in the event record rather than being allowed to decay into their ground-state daughters. All other unstable particles are allowed to decay only if their nominal proper decay length satisfies ($c\tau_0 < 0.01$~mm). The resulting event record thus contains the heavy-flavour hadrons prior to their decays, together with the accompanying light-particle final state.

\subsection{Hadronisation tunes}
\label{sec:hadronisation-tunes}

The Monash tune is used as the baseline configuration. It employs conventional Lund string fragmentation, supplemented by the Bowler modification to account for heavy-quark fragmentation, and uses \texttt{ColourReconnection:mode = 0} (CR mode~0), corresponding to the MPI-based colour-reconnection model. Its fragmentation parameters were tuned primarily to \(\mathrm{e}^+\mathrm{e}^-\) data from LEP and SLD, while the parameters governing initial-state radiation and MPIs were constrained using measurements from hadron collisions, including pp data from the LHC. Both SJ and CP are based on the Monash 2013 tune~\cite{Skands:2014pea}, but incorporate additional modelling ingredients and configuration-specific parameter settings. The comparison therefore involves complete tune configurations, rather than isolated modifications of individual hadronisation mechanisms. For SJ, we use the original Mode-2 parameter set of Christiansen and Skands~\cite{Christiansen:2015yqa}, while for CP we use the T1 parameter set of Altmann et al.~\cite{Altmann:2025afh}.

The main characteristic feature of SJ is its use of a QCD-inspired colour-reconnection model, implemented through \texttt{ColourReconnection:mode = 1} (CR mode~1), which allows the formation of junction colour topologies. The model combines a string-length minimisation criterion with a causal prescription that accounts for time-dilation effects, controlled by \texttt{ColourReconnection:timeDilationMode = 2}. The resulting three-leg junction configurations provide an additional mechanism for baryon formation, complementing the conventional diquark-based baryon production in string fragmentation. This provides a natural motivation for investigating whether SJ enhances the baryonic contribution to the heavy-flavour balancing yields relative to the Monash baseline.

The CP tune is a more recent development, designed to describe, among other observables, the enhanced production of strange and baryonic hadrons observed in high-multiplicity pp collisions at the LHC \cite{Altmann:2025afh,ALICE:2016fzo,ALICE:2020wla,ALICE:2023wbx,ALICE:2020wfu,ALICE:2017thy,ALICE:2021rzj,ALICE:2021psx,ALICE:2021bli,ALICE:2017dja,ALICE:2022cop,LHCb:2023wbo,CMS:2016zzh,ATLAS:2024nbm}. Like SJ, it employs CR mode~1 and  allows, therefore, the formation of junction topologies. In addition, it incorporates local string-density effects, estimated from the rapidity overlap of the strings. In regions of higher string density, these effects modify the effective string tension and the transverse-momentum distribution generated during string breaking. The tune also includes a dedicated modification of strangeness production at junctions and a suppression of diquark production. These ingredients introduce an additional dependence of hadron production on the local string environment, potentially enhancing the relative production of strange and baryonic hadrons in events with a high density of overlapping strings. The comparison with SJ therefore allows us to investigate how these additional close-packing effects influence the species composition of heavy-flavour balancing.

For this analysis, the published tune configurations are used without further retuning to heavy-flavour data, providing a comparison based on the respective model predictions. Table~\ref{tab:pythia_parameters} summarises the parameter settings that differ among the three configurations. A more detailed overview of the colour-reconnection parameters is provided in Table~\ref{tab:additional-hadronisation-settings} in Appendix~\ref{sec:additional-hadronisation-settings}.

\begin{table}[htbp]
    \centering
    \caption{Selected PYTHIA 8.317 parameter settings for the Monash,
    SJ and CP configurations. N/A denotes parameters
    not applicable to the corresponding colour-reconnection model.}
    \label{tab:pythia_parameters}

    \small
    \setlength{\tabcolsep}{4pt}
    \renewcommand{\arraystretch}{1.05}

    \begin{tabularx}{\textwidth}{@{}
        >{\raggedright\arraybackslash}X
        >{\centering\arraybackslash}p{0.19\textwidth}
        >{\centering\arraybackslash}p{0.20\textwidth}
        >{\centering\arraybackslash}p{0.19\textwidth}
        @{}
    }
        \toprule
        \textbf{Parameter}
        & \textbf{Monash}
        & \textbf{SJ}
        & \textbf{CP} \\
        \midrule

        \texttt{ColourReconnection:mode}
        & 0 & 1 & 1 \\

        \texttt{BeamRemnants:remnantMode}
        & 0 & 1 & 1 \\

        \texttt{StringZ:aLund}
        & 0.68 & 0.36 & 0.68 \\

        \texttt{StringZ:bLund} [GeV$^{-2}$]
        & 0.98 & 0.56 & 0.98 \\

        \texttt{StringZ:useOldAExtra}
        & on & on & off \\

        \texttt{StringFlav:probStoUD}
        & 0.217 & 0.200 & 0.217 \\

        \texttt{StringFlav:probQQtoQ}
        & 0.081 & 0.078 & 0.081 \\

        \texttt{StringFlav:probQQ1toQQ0join}
        & $\{0.5,\,0.7,\,0.9,\,1.0\}$
        & $\{0.0275,\,0.0275,$ \newline $0.0275,\,0.0275\}$
        & $\{0.5,\,0.7,\,0.9,\,1.0\}$ \\

        MPI \texttt{pT0Ref} [GeV] (7 TeV ref.)
        & 2.280 & 2.150 & 2.194 \\

        CR \texttt{mPseudo} [GeV]
        & N/A & 0.300 & 0.403 \\

        CR \texttt{m0} [GeV]
        & N/A & 0.300 & 0.618 \\

        CR \texttt{junctionCorrection}
        & N/A & 1.20 & 1.349 \\

        \texttt{ClosePacking:doClosePacking}
        & off & off & on \\

        \bottomrule
    \end{tabularx}
\end{table}

\subsection{Event activity and multiplicity intervals}
\label{sec:event-activity-multiplicity}

To investigate how heavy-flavour balancing correlations depend on the event environment, each simulated sample is divided into intervals according to the charged-particle multiplicity, $N_{\mathrm{ch}}$, which is used as an estimator of event activity. We define $N_{\mathrm{ch}}$ as the number of final-state charged particles that do not contain charm or beauty quarks, satisfying $p_{\mathrm{T}} > 0.15$~GeV/$c$ and $|\eta|\leq4$. The particle-level definition follows the decay convention described above. Excluding heavy-flavour hadrons from the estimator prevents the particles whose balancing correlations are studied from contributing directly to the event-activity measure.

The three configurations predict different charged-particle multiplicity distributions. We therefore determine the percentile boundaries independently for each tune, such that a given interval corresponds to the same relative activity range within each generated sample. Using common numerical intervals in $N_{\mathrm{ch}}$ would instead select different fractions of the samples. Consequently, the same multiplicity-class label in different tunes does not imply equal absolute multiplicities, equal underlying MPI activity, or identical event environments. Since $N_{\mathrm{ch}}$ is integer-valued, events with the same multiplicity are assigned to the same interval. The resulting numerical boundaries are listed in Table~\ref{tab:multiplicity_classes} in Appendix~\ref{sec:additional-hadronisation-settings}.

\begin{figure}
    \centering
    \includegraphics[width=0.97\linewidth]{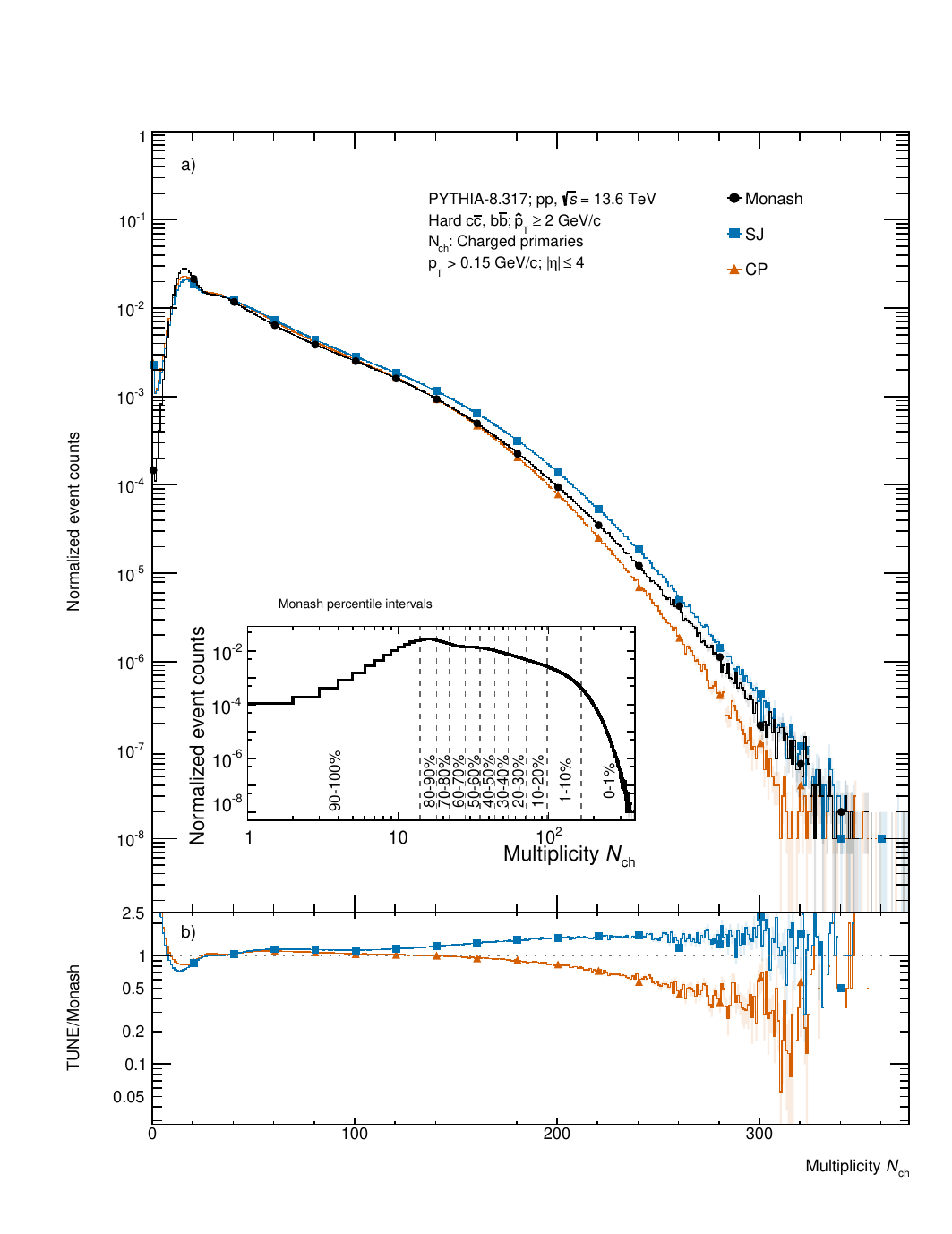}
    \caption{ (upper) Separately normalised charged-particle multiplicity distributions for the Monash, SJ and CP samples. The inset illustrates the multiplicity percentile ranges using the Monash distribution, with dashed lines marking the interval boundaries. Multiplicity interval boundaries are determined separately for each tune, as described in Section~\ref{sec:event-activity-multiplicity}. (lower) Ratios of the SJ and CP multiplicity distributions with respect to the Monash tune.}
    \label{fig:multiplicity_distribution}
\end{figure}

Figure~\ref{fig:multiplicity_distribution} shows the normalised charged-particle multiplicity distributions for the three tunes. The multiplicity intervals are defined in percentiles of each distribution, with $0$--$1\%$ denoting the highest- and $90$--$100\%$ the lowest-multiplicity events.

\subsection{Heavy-flavour hadron balancing}
\label{sec:pair-selection}

In the present analysis, a trigger is defined as a final heavy hadron of a given species $t$, within $|\eta|\leq4$, which establishes a common, broad but finite acceptance. We compare representative ground-state meson and baryon triggers in both the charm and beauty sectors to investigate whether the associated heavy-hadron species-composition depends on the trigger species. For this study, we chose to focus on the lightest meson and baryon in each sector as trigger species: $D^{0}$ and $\Lambda_{c}^{+}$ for charm, and $B^{+}$ and $\Lambda_{b}^{0}$ for beauty. Every particle of these species is accepted as a trigger, independent of their origin---i.e. created in a hard scattering, showering, or any other production mechanism.

Each trigger is paired with every distinct eligible associated heavy hadron in the same event. Associates are identified open-heavy hadrons produced directly in hadronisation, selected through PYTHIA status codes 81--89 and required to satisfy $|\eta|\leq4$; a particle cannot be paired with itself. These pseudorapidity requirements can lead to events having eligible triggers with no accepted associate, so the trigger normalisation includes all eligible triggers regardless of whether they have an accepted associate or not. Both selections are inclusive in transverse momentum, so no fixed hadron-level $p_{\mathrm{T}}$ threshold or ordering between trigger and associate $p_{\mathrm{T}}$ is imposed. Through this inclusive transverse-momentum selection we retain the soft heavy-hadron population.

For a fixed generator configuration and multiplicity interval $m$, opposite-sign (OS) and same-sign (SS) describe the signs of the relevant heavy-flavour quantum numbers, rather than electric charge. For a specified OS associate species $a$, the SS reference is its charge-conjugate species $\bar a$, and the trigger species remains unchanged between the two samples. Pairs are counted as ordered trigger--associate combinations; thus, an event can contribute multiple triggers and multiple pairs. 

Following Eqs.~\eqref{eq:charge-balance} and \eqref{eq:correlation-function}, we construct the OS and SS per-trigger pair densities for each trigger species, associate species and multiplicity interval. We project these densities onto the azimuthal separation $\Delta\varphi$ by summing over all accepted pseudorapidity separations, retaining the requirement of $|\eta|\leq 4$ for both particles. In each angular bin, the pair count is divided by the number of selected triggers and the bin width. The difference between the OS and SS densities is the differential balancing yield, $Y^{(m)}_{t,a}(\Delta\varphi)$. It measures the excess of OS associates over the SS per trigger per radian. This subtraction removes background contributions that are equal in the two distributions and shows how the excess of opposite-heavy-flavour associates is distributed in azimuth.

 To compare the associate species without restricting the pairs to a particular angular region, we integrate the differential balancing yield over the complete azimuthal interval according to:

 \begin{equation}
    Y_{t,a}^{(m)}
    =
    \int_{-\pi/2}^{3\pi/2}
    Y_{t,a}^{(m)}(\Delta\varphi)\,
    d\Delta\varphi .
    \label{eq:integrated-balancing-yield}
\end{equation}

The integrated balancing yield is therefore the total OS pair count minus the total SS pair count, divided by the number of selected triggers in the same multiplicity interval. The differential yield describes how this excess of OS over SS is distributed in azimuth, while the integrated yield measures the net yield of the specified associate species for a given trigger. Comparing the integrated yields across associate species shows whether the net balancing yield is carried mainly by mesons or baryons, and how this composition changes with the trigger species and hadronisation model.

In our analysis, central values are calculated from pooled counts, and statistical uncertainties are estimated with a delete-one-block jackknife using ten blocks of ten million events per tune, as described in Appendix~\ref{sec:stat-uncertainty}. After each deletion, we recompute the observable, including its normalisation, ratios and affected multiplicity boundaries. This preserves correlations between OS and SS counts and between quantities sharing a denominator. Through this set of observables, we are able to test how the angular structure and species composition of heavy-flavour balancing vary between the three generator configurations and multiplicity intervals.

%% file: sections/Results.tex
\section{Results}
\label{Sec:Results}

\begin{figure}
    \centering
    \includegraphics[width=1\linewidth]{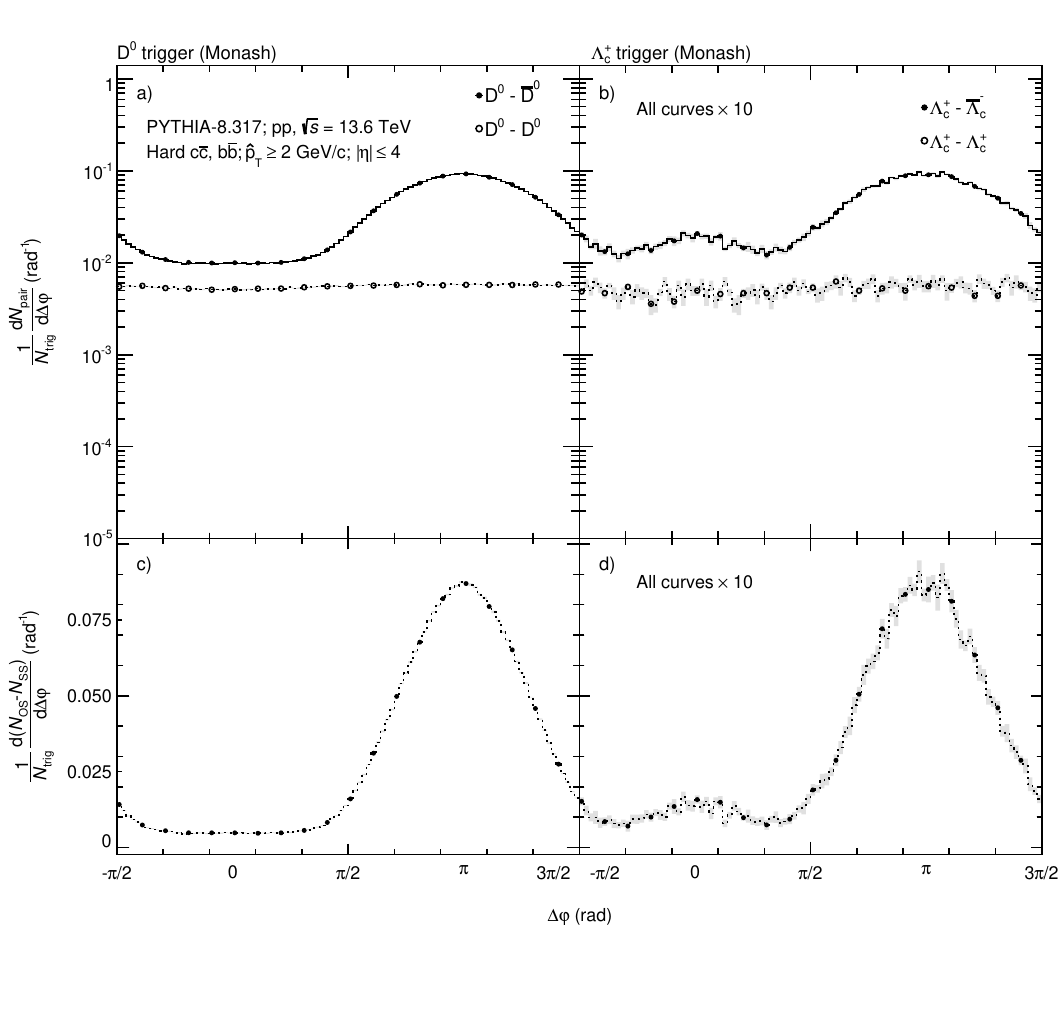}
    \caption{Two-particle correlations of $D^{0}$ mesons (left) and $\Lambda_{c}^+$ baryons (right) with the PYTHIA-8 Monash tune. Note that the results for $\Lambda_{c}^+$ baryons are scaled up by a factor of 10 to bring them to the same level as the relevant results for mesons. The opposite- and same-charm correlations are shown (top), together with their difference (bottom), which cancels contributions equal in the two distributions. A distinct away-side peak is identified around $\Delta \varphi = \pi$.}
    \label{fig:correlations_charm}
\end{figure}

The upper panels of figure~\ref{fig:correlations_charm} show the azimuthal correlation functions, i.e. the projections in $\Delta \varphi$ of the expression defined in Eq.~\eqref{eq:correlation-function}, for the lightest charm mesons and baryons, integrated over event multiplicity in simulated pp collisions at $\sqrt{s}=13.6$~TeV using the Monash tune of PYTHIA-8. The left panel shows correlations with a $D^0$ trigger, comparing $D^0$--$\overline{D}^0$ pairs (filled markers) with $D^0$--$D^0$ pairs (open markers), while the right panel presents the corresponding distributions for $\Lambda_c^+$ triggers correlated with $\overline\Lambda_c^-$ and $\Lambda_c^+$ hadrons, respectively. Note that the latter set of data points is scaled up by a factor of 10 to bring them to the same level as their meson counterparts. In both cases, the trigger particles are required to satisfy $|\eta|\leq4$. For both meson and baryon triggers, the opposite-charm correlations exhibit a pronounced peak around $\Delta\varphi=\pi$, characteristic of the away side. This structure is consistent with the predominantly back-to-back production of $c\bar{c}$ pairs in hard partonic scatterings. Although this initial correlation can be broadened by parton showers, higher-order radiation, additional partonic interactions and hadronisation, it remains visible in the angular distributions of the resulting charm hadrons. On the other hand, no pronounced enhancement is observed around $\Delta\varphi=0$, i.e. the near side, except for a hint of a near-side enhancement of $\overline\Lambda_c^-$ and $\Lambda_c^+$. This is consistent with the expectation that the two heavy quarks produced in the same hard scattering event are generally separated in azimuth, rather than being emitted collinearly. The near-side region can nevertheless receive contributions from higher-order production mechanisms, parton showers and decays; in particular, non-prompt charm hadrons originating from beauty-hadron decays can, in principle, introduce additional angular correlations, but they are disabled in this work. The distributions away from the near- and away-side regions contain a mixture of residual heavy-flavour correlations and contributions from other event activity. Finally, the same-charm correlation functions are comparatively flat as a function of $\Delta\varphi$, with no prominent away-side structure. Within this sample and for these selections, this indicates that the same-charm combinations do not exhibit a strong angular correlation comparable to that seen for opposite-charm pairs. The contrast between the two distributions provides the basis for the subtraction used below to isolate the component associated with heavy-flavour balancing.

To further isolate the angular correlations associated with heavy-flavour balancing, the lower panels of Fig.~\ref{fig:correlations_charm} show the difference between the opposite- and same-charm correlation functions. Since charm is produced predominantly in $c\overline{c}$ pairs, the charm quantum number carried by a trigger hadron must be compensated by the remaining final state. Correlations with hadrons carrying the opposite charm quantum number therefore contain the contribution from the balancing heavy quark, whereas same-charm correlations estimate contributions not directly associated with this balancing. Subtracting the same-charm distribution from the opposite-charm distribution suppresses contributions common to both samples and makes the characteristic away-side structure more apparent. For both $D^0$ and $\Lambda_{c}^+$ triggers, the resulting distributions are consistent with a balancing yield concentrated around $\Delta\varphi=\pi$, while remaining close to zero around the near side. As noted above, the subtracted correlations for baryons provide a hint for a near-side structure which is, however, not significant within the current level of uncertainties. This behaviour supports the interpretation that the measured angular structure is primarily driven by the production and subsequent hadronisation of the heavy-quark pair. However, the subtraction does not completely remove all uncorrelated or background contributions: same-charm pairs can themselves arise from correlated production mechanisms, and opposite-charm pairs can receive contributions beyond the balancing of the original heavy-quark pair. The resulting distribution is therefore best understood as a differential measure of the excess of opposite-charm correlation relative to the same-charm reference, providing a more relevant probe of heavy-flavour balancing than either correlation function alone.

%======================================%

\begin{figure}
    \centering
    \includegraphics[width=1\linewidth]{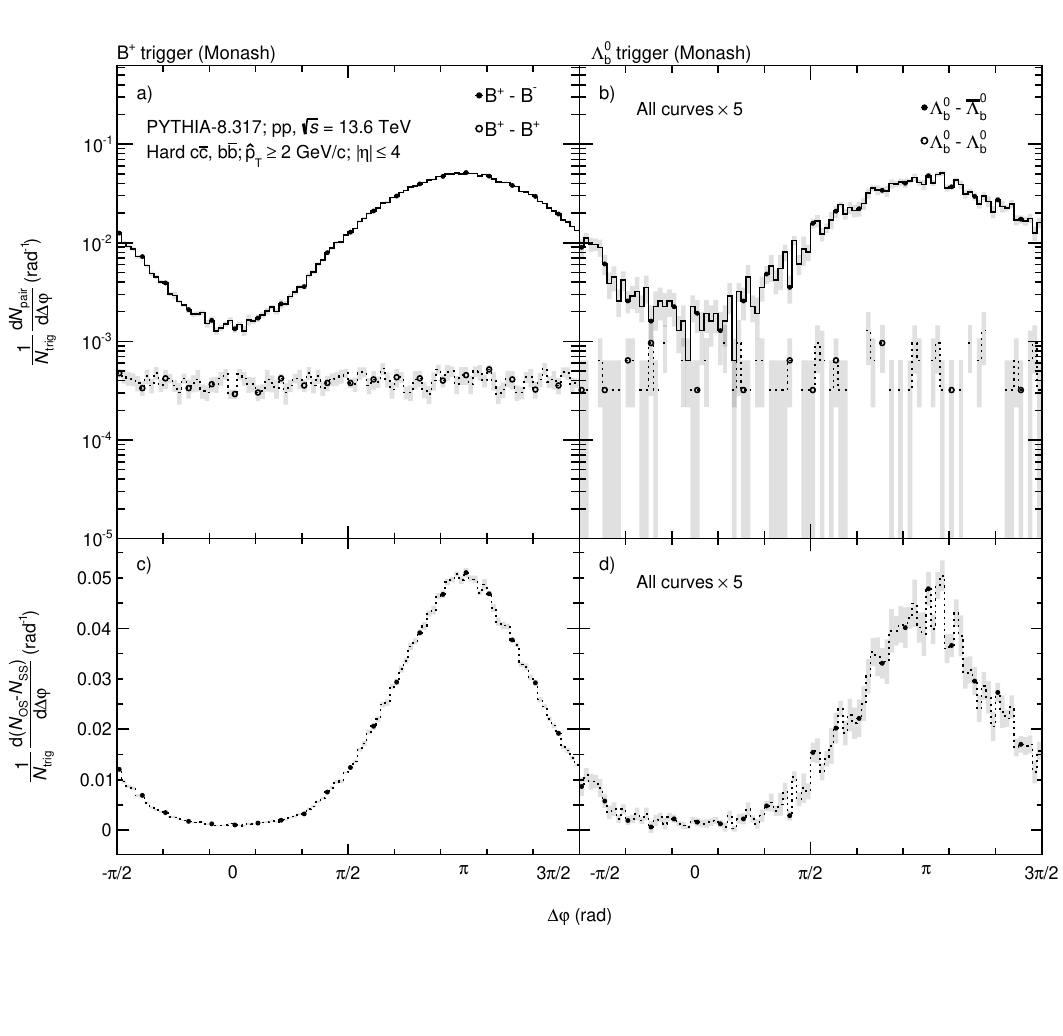}
    \caption{Two-particle correlations of $B^{+}$ mesons (left) and $\Lambda_{b}^0$ baryons (right) with the PYTHIA-8 Monash tune. Note that the results for $\Lambda_{b}^0$ baryons are scaled up by a factor of 5 to bring them to the same level as the relevant results for mesons. The opposite- and same-beauty correlations are shown (top), together with their difference (bottom), which cancels contributions equal in the two distributions. A distinct away-side peak is identified around $\Delta \varphi = \pi$.}
    \label{fig:correlation_beauty}
\end{figure}

Figure~\ref{fig:correlation_beauty} shows the corresponding azimuthal correlation functions for beauty hadrons produced in pp collisions at $\sqrt{s}=13.6$~TeV using the Monash tune of PYTHIA-8. In panel (a), $B^+$ mesons are selected as trigger particles and correlated with $B^-$ mesons (filled circles) and $B^+$ mesons (open circles). Panel (b) presents the corresponding correlations for $\Lambda_{b}^0$ triggers paired with $\overline{\Lambda}_{b}^0$ (filled markers) and $\Lambda_{b}^0$ (open markers). Note that the results for $\Lambda_{b}^0$ baryons are scaled up by a factor of 5 to bring them to the same level as the relevant results for mesons. These panels exhibit the same qualitative features as observed in the charm sector: correlations between hadrons carrying opposite beauty quantum numbers show a pronounced away-side structure around $\Delta\varphi=\pi$, consistent with the production of $b\overline{b}$ pairs in hard partonic scatterings, whereas the same-beauty correlations remain comparatively flat within the uncertainties. The persistence of this pattern for both meson and baryon triggers indicates that the initial angular correlation between the heavy quark and antiquark remains visible after hadronisation, despite the different hadronic species formed. As in the charm sector, no pronounced near-side enhancement is observed in the opposite-beauty correlations. 

The lower panels, (c) and (d), show the corresponding difference between opposite- and same-beauty correlation functions. The resulting balancing distributions are concentrated on the away side for both trigger species, supporting the interpretation that the dominant angular structure reflects the balancing of the initially produced beauty quantum number. A notable feature is that the balancing yield is larger for $B^+$ triggers than for $\Lambda_{b}^0$ triggers. This difference suggests that, within the Monash tune and for the selected species and kinematic range, the opposite beauty quantum number is more frequently observed in the measured beauty-meson channel than in the corresponding beauty-baryon channel. This observation is consistent with the relative suppression of baryon formation in the default fragmentation model, although the correlation functions alone do not establish whether the difference arises solely from the relative production probabilities of the trigger and associated species. The species-dependent balancing yields discussed below provide a more direct way to investigate how the hadronisation model distributes the balancing beauty quantum number among mesons and baryons.

\begin{figure}
    \centering
    \includegraphics[width=1\linewidth]{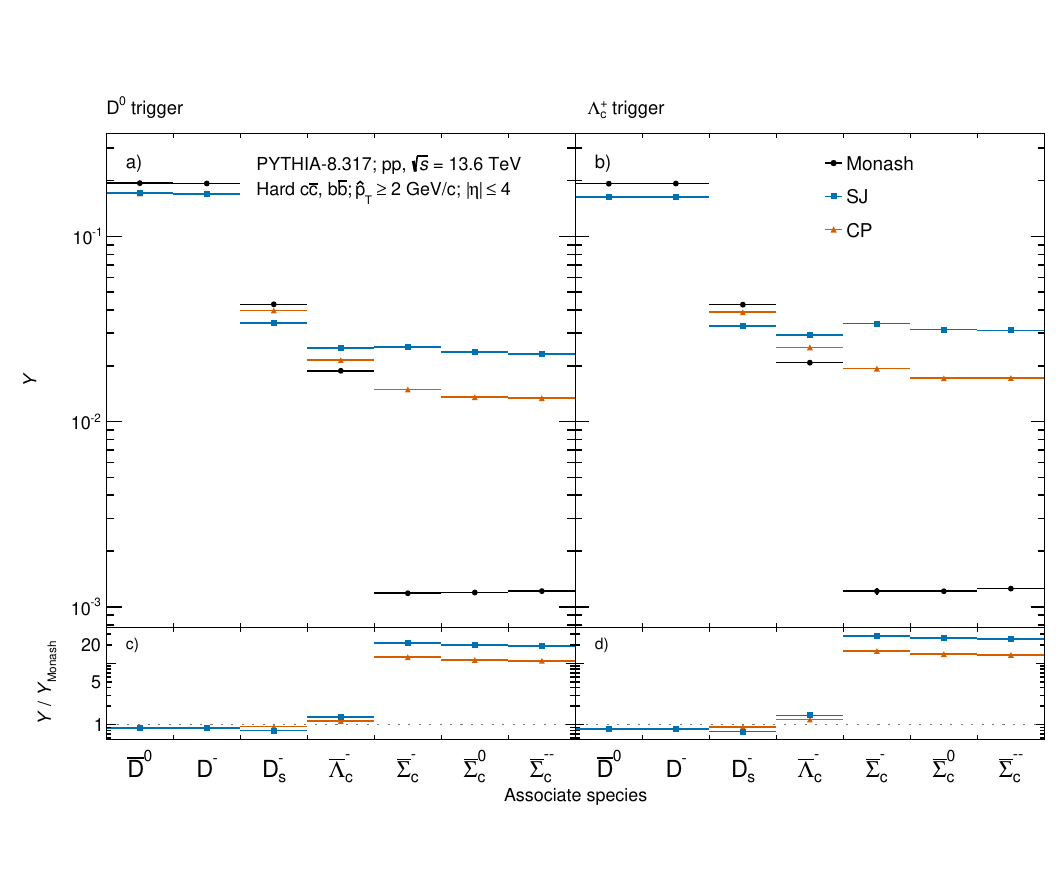}
    \caption{Balancing yield for $D^{0}$ trigger (left) and $\Lambda_{c}^+$ trigger (right) with the Monash, SJ and CP tunes. Ratios with respect to the Monash tune are shown in the lower panels. An increase of balancing yield is seen for the baryons when the SJ and CP tunes are used. This increase is particularly strong for the isospin-1 particles.}
    \label{fig:balancing_integrated_charm}
\end{figure}

Figure~\ref{fig:balancing_integrated_charm} shows the integrated balancing yields for charm-hadron correlations, evaluated over the full multiplicity range, with $D^0$ and $\Lambda_{c}^+$ triggers in the left and right panels, respectively. For each particle species, the yield is obtained by integrating the balancing distributions of the type shown in the lower panels of Fig.~\ref{fig:correlations_charm} over $\Delta\varphi$. Results are compared for the Monash, SJ and CP tunes of PYTHIA-8, allowing us to investigate how the hadronisation model affects the species composition of the balancing charm quantum number. All three tunes exhibit a common hierarchy among the dominant mesonic balancing partners: $\overline{D}^0$ and $D^-$ contribute the largest yields, followed by $D_{\mathrm{s}}^-$. This pattern indicates that the balancing charm quantum number is predominantly carried by the lightest charm-meson states included in the measurement. More pronounced differences are observed in the baryon sector. In the Monash configuration, the baryonic balancing yields are comparatively small, with the $\Sigma_{c}$ family particularly strongly suppressed. In contrast, both SJ and CP produce substantially larger balancing yields for charm baryons, irrespective of whether the trigger is a $D^0$ meson or a $\Lambda_{c}^+$ baryon. This is compensated by lower yields for the principal mesonic balancing partners, indicating a redistribution of the balancing yield among hadron species rather than a uniform increase across all channels. The change is consistent with the enhanced baryon production associated with the alternative hadronisation models. 
In the SJ configuration, the junction colour topologies provide an additional mechanism for baryon formation beyond conventional diquark-based string fragmentation. In addition, CP also enhances baryon production through modifications to hadronisation in a dense string environment, although the magnitude of the enhancement is generally smaller than in SJ. The most striking difference concerns the $\Sigma_{c}$ family, whose balancing yield increases by approximately ten to thirty times relative to Monash in both the SJ and CP tunes, for both trigger species. This enhancement is substantially larger than the corresponding change in the $\Lambda_{c}$ balancing yield, suggesting that the model differences extend beyond the overall baryon fraction to the relative production of specific baryon states. One possible explanation involves differences in the production of diquark configurations and their subsequent fragmentation into baryons. The $\Lambda_{c}$ ground state has isospin $I=0$, whereas the $\Sigma_{c}$ is an isospin triplet with $I=1$. The enhanced $\Sigma_{c}$ yield may therefore be sensitive to the relative probabilities assigned to the relevant diquark configurations and to the parameters controlling their formation in PYTHIA-8. This interpretation requires further verification against the tune settings, in particular with the parameter \texttt{StringFlav:probQQ1toQQ0join}, listed in Table~\ref{tab:pythia_parameters}. Overall, these results demonstrate that the hadronisation model affects not only the total abundance of charm baryons but also the species composition of the particles carrying the balancing charm quantum number.

%=================================================================%

\begin{figure}
    \centering
    \includegraphics[width=1\linewidth]{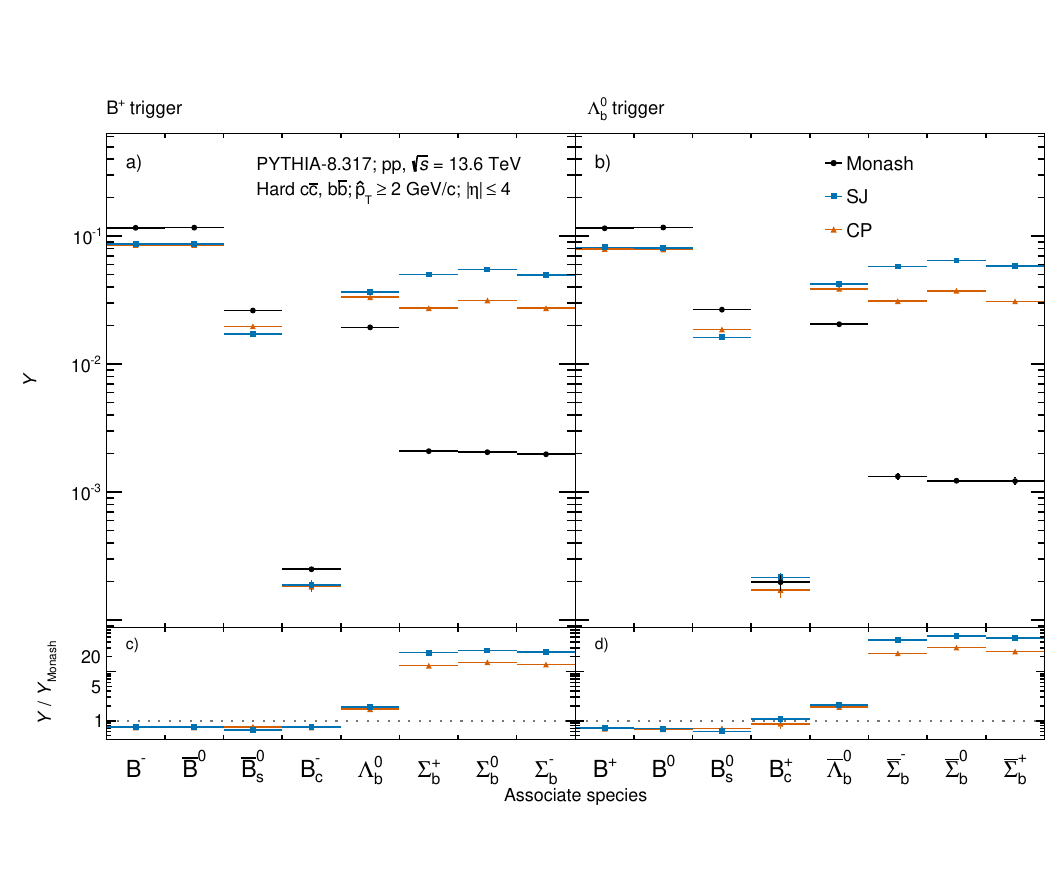}
    \caption{Balancing yield for $B^{+}$ trigger (left) and $\Lambda_{b}^0$ trigger (right) with the Monash, SJ and CP tunes. Ratios with respect to the Monash tune are shown in the lower panels. For the SJ and CP tunes we observe an increase of balancing yield for the baryons when the SJ and CP tunes are used. This increase is particularly strong for the isospin-1 particles.}
    \label{fig:balacing_integrated_beauty}
\end{figure}

Figure~\ref{fig:balacing_integrated_beauty} presents the integrated balancing yields, evaluated over the full multiplicity range, for beauty hadrons in pp collisions at $\sqrt{s}=13.6$~TeV, comparing the Monash, SJ and CP tunes of PYTHIA-8. The left and right panels show the results for $B^+$ and $\Lambda_{b}^0$ triggers, respectively, correlated with various beauty mesons and baryons carrying the opposite beauty quantum number. As in the charm sector, all three tunes exhibit a clear preference for the lightest beauty-meson balancing partners, with $B^-$ and $\overline{B}^0$ providing the largest contributions. This common hierarchy suggests that, despite the differences between the hadronisation models, the dominant balancing channels remain associated with the production of the lightest accessible beauty mesons. The differences between the tunes, however, become more apparent among the less abundant balancing species. In Monash, the $\overline{B}_{\mathrm{s}}^0$ contribution remains larger than the contributions from the beauty-baryon channels, whereas in both SJ and CP the baryonic balancing yields become more prominent, exceeding the $\overline{B}_{\mathrm{s}}^0$ contribution. This pattern is qualitatively consistent with the charm results, where the alternative tunes enhance baryonic balancing at the expense of the dominant mesonic channels. It indicates that the impact of these models extends to beauty hadronisation and is not restricted to charm. In both the SJ and CP tunes, the enhancement is particularly pronounced for baryons belonging to the $\Sigma_{b}$ family, whose balancing yields are substantially larger than in the Monash tune. The corresponding behaviour in the charm and beauty sectors points to a common sensitivity to the mechanisms governing baryon formation, including junction topologies and hadronisation in a dense string environment. At the same time, the relative contributions of the individual balancing species differ between charm and beauty. These differences may reflect the distinct fragmentation probabilities and available hadron species in the two sectors, and motivate a more direct comparison of the baryon-to-meson balancing ratios as a function of event activity.

\begin{figure}
    \centering
    \includegraphics[width=0.84\linewidth]{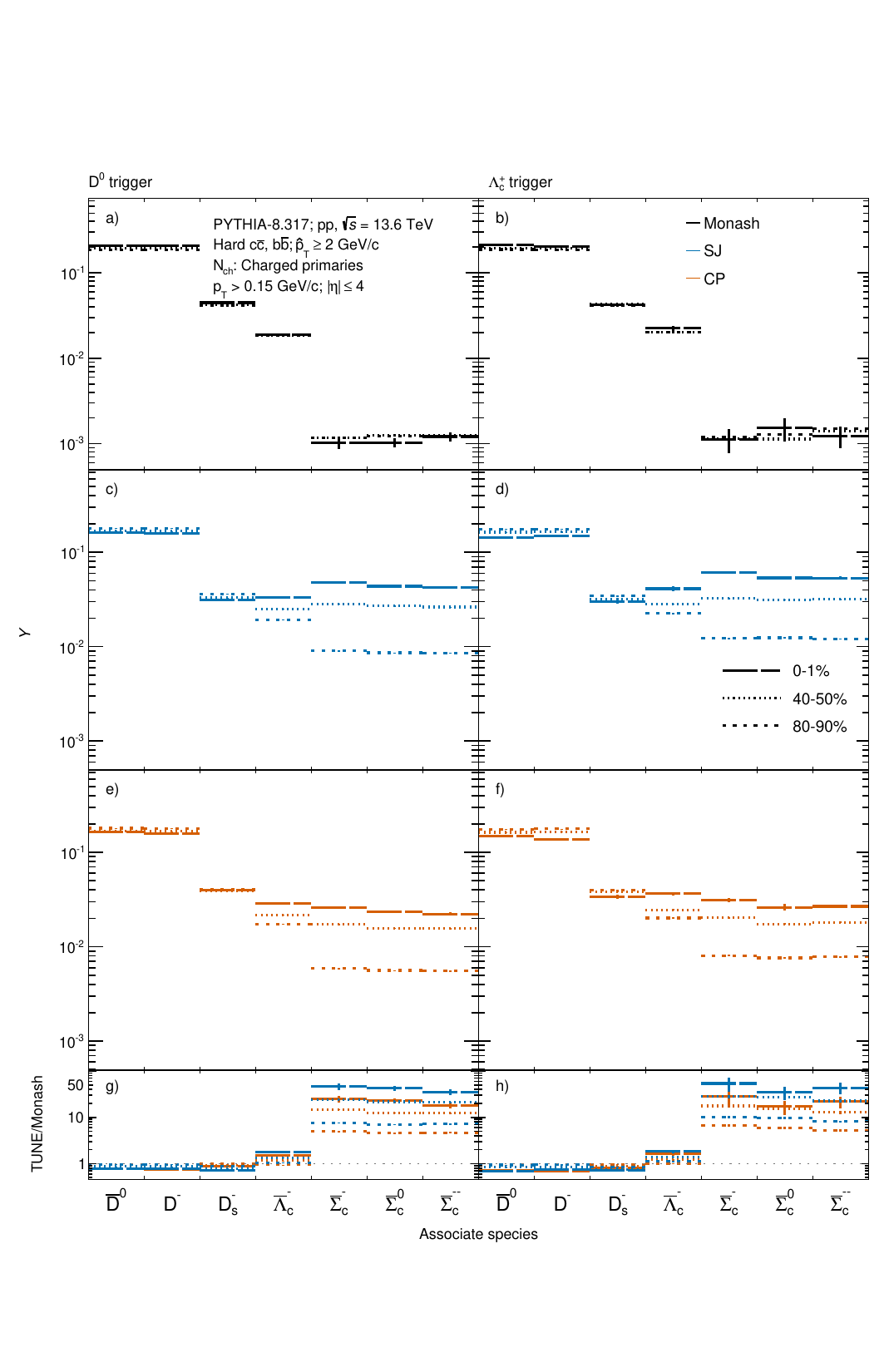}
    \caption{Balancing yield for $D^{0}$ trigger (left) and $\Lambda_{c}^+$ trigger (right) with the Monash, SJ and CP tunes for three event multiplicity classes: 0-1\%, 40-50\% and 80-90\%, as defined in Table~\ref{tab:multiplicity_classes}. Ratios with respect to the Monash tune are shown in the lower panels. An increase of balancing yield is seen for the baryons when the SJ and CP tunes are used and the ratio gets larger with increasing event multiplicity. This effect is particularly strong for the isospin-1 particles.}
    \label{fig:supplemental_balancing_activity_charm_extremes}
\end{figure}

\begin{figure}
    \centering
    \includegraphics[width=0.84\linewidth]{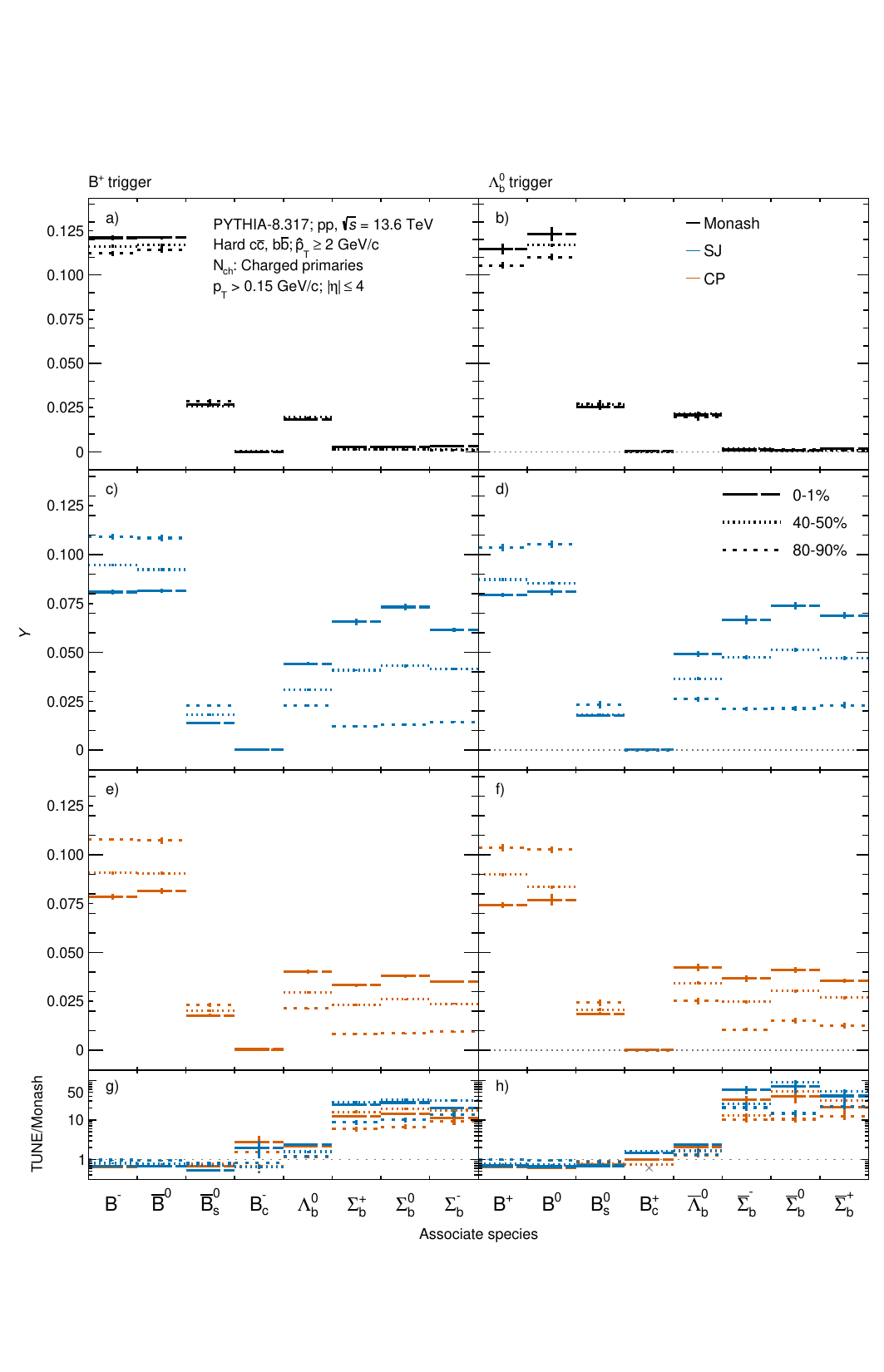}
    \caption{Balancing yield for $B^{+}$ trigger (left) and $\Lambda_{b}^0$ trigger(right) with the Monash, SJ and CP tunes for three event multiplicity classes: 0-1\%, 40-50\% and 80-90\%, as defined in Table~\ref{tab:multiplicity_classes}. Ratios with respect to the Monash tune are shown in the lower panels. An increase of balancing yield is seen for the baryons when the SJ and CP tunes are used and the ratio gets larger with increasing event multiplicity. This effect is particularly strong for the isospin-1 particles.}
    \label{fig:supplemental_balancing_activity_beauty_extremes}
\end{figure}

The multiplicity-dependent integrated balancing spectra for charm and beauty are shown in Figs.~\ref{fig:supplemental_balancing_activity_charm_extremes} and \ref{fig:supplemental_balancing_activity_beauty_extremes}, respectively. The results are presented for three event-activity intervals in pp collisions at \(\sqrt{s}=13.6\)~TeV: the highest-multiplicity 1\%, the 40--50\% interval, and the 80--90\% interval. The corresponding results are represented by long-dashed, dotted, and short-dashed lines, respectively. In each figure, the upper, middle, and lower rows correspond to the Monash, SJ, and CP tunes, while the left and right columns show the results for \(D^0\) and \(\Lambda_{c}^+\) triggers in the charm sector, and for \(B^+\) and \(\Lambda_{b}^0\) triggers in the beauty sector. In both sectors, the Monash results exhibit little to only a modest dependence on event multiplicity, consistent with the relatively stable species composition observed in the multiplicity-integrated yields. In contrast, SJ and CP show a clear evolution with increasing event activity: the balancing yields associated with mesonic channels generally decrease, while those associated with baryonic channels increase. This pattern is consistent with a multiplicity-dependent redistribution of the balancing heavy-flavour yield towards baryons in the alternative hadronisation models. The effect is particularly pronounced in the beauty sector and for baryonic balancing channels, indicating that the response to event activity depends both on the hadronisation model and on the heavy-flavour species.

%=================================================================%

\begin{figure}
    \centering
    \includegraphics[width=1\linewidth]{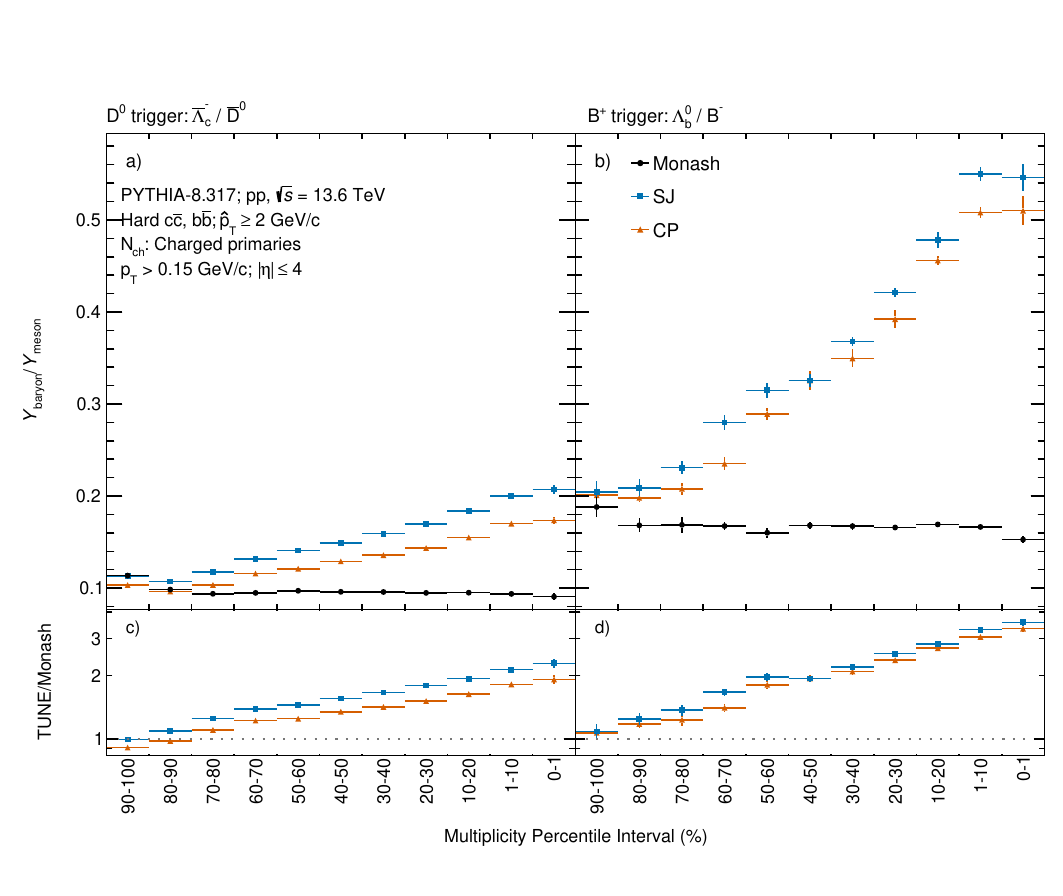}
    \caption{Baryon-to-meson ratio for $\overline{\Lambda}_{c}^-$/$\overline{D}^{0}$ (left) and $\Lambda_{b}^0$/$B^{-}$ (right) as a function of the event multiplicity, generated with the Monash, SJ and CP tunes. Ratios with respect to the Monash tune are shown in the lower panels. The plots show an increasing trend of the baryon-to-meson ratio as a function of multiplicity, both for the charm and for the beauty cases. In both cases the ratio saturates in the largest multiplicity classes.}
    \label{fig:balancing_baryon_meson_activity}
\end{figure}

Figure~\ref{fig:balancing_baryon_meson_activity} shows the ratio of baryonic to mesonic balancing yields for \(D^0\) and \(B^+\) triggers in the left and right panels, respectively, as a function of event multiplicity in pp collisions at \(\sqrt{s}=13.6\)~TeV for the three PYTHIA-8 tunes considered. This ratio provides a differential probe of the different hadronisation mechanisms, testing how colour reconnection, junction formation and hadronisation in a dense string environment affect the relative contributions of baryonic and mesonic channels to heavy-flavour balancing. For Monash, represented by black circles, the ratio exhibits no strong multiplicity dependence, although a modest decrease may be present in the highest-activity events. This suggests that, within the default string-fragmentation model, the relative contributions of the selected baryonic and mesonic balancing channels remain approximately stable across the event-activity classes considered. The ratio is nevertheless substantially larger for beauty than for charm, indicating a stronger relative contribution from the selected baryonic balancing channel in the beauty sector for this tune.

In contrast, both SJ and CP exhibit a pronounced increase in the baryon-to-meson balancing ratio with event multiplicity. In the highest-multiplicity 1\% of events, the ratio reaches approximately 0.2 for charm and exceeds 0.5 for beauty. As shown in the lower panels of fig.~\ref{fig:balancing_baryon_meson_activity}, these values are approximately two and three times larger, respectively, than those obtained with Monash. This behaviour is consistent with the enhanced baryon production associated with the alternative hadronisation models. In SJ, additional colour topologies involving junctions provide a mechanism for baryon formation beyond conventional diquark-based string fragmentation. In turn, CP modifies hadronisation in a dense string environment, where interactions among nearby strings can affect the resulting hadron species. The increasing ratios suggest that the impact of these mechanisms on heavy-flavour balancing becomes more pronounced in high-multiplicity events, in line with the multiplicity-dependent changes in the balancing spectra discussed above. Thus, the results connect the event-activity dependence of heavy-flavour correlations to the broader question of how hadronisation mechanisms can modify baryon production in high-density pp collisions.

This multiplicity dependence can be understood in the context of MPIs and the resulting colour structure of pp collisions. As the event activity increases, additional partonic scatterings and parton-shower emissions contribute to a more complex environment of colour connections, which can be rearranged by colour reconnection before hadronisation. In the SJ model, such reconnections can generate junction topologies that provide additional mechanisms for baryon formation beyond conventional string fragmentation. The increasing baryon-to-meson balancing ratio is therefore consistent with a growing contribution from these mechanisms in high-activity events. In CP, the trend may instead be associated with modifications to hadronisation in a dense environment of overlapping strings. These interpretations are qualitatively consistent with studies of baryon production and colour reconnection in PYTHIA-8~\cite{Lonnblad:2023kft}, as well as recent investigations of string close packing~\cite{Altmann:2025afh}. However, event multiplicity is only an indirect indicator of the underlying MPI activity and string density, and the present results do not isolate their individual contributions. The observed increase should therefore be interpreted as evidence of a stronger baryonic contribution to heavy-flavour balancing in these models, rather than as a direct measurement of MPI-driven baryon production.

Finally, the ratio remains larger in the beauty than in the charm sector, exceeding the charm value by more than a factor of 1.7 in the highest-multiplicity events. This difference highlights the distinct quantitative behaviour of charm and beauty hadronisation in the models studied. While the increase with multiplicity is consistent with a stronger baryonic contribution in the alternative tunes, the ratio alone does not establish which specific mechanism is responsible, nor whether the charm--beauty difference originates from the heavy-quark mass, flavour-dependent fragmentation probabilities or the particular species included in the observable. Taken together, these results show that the relative baryonic contribution to heavy-flavour balancing depends on both the event environment and the hadronisation model, making these correlations a useful complementary study to inclusive baryon-to-meson yield measurements.

%% file: sections/Summary.tex
\section{Conclusion}
\label{Sec:Summary}
In this article, we presented a systematic generator-level study of heavy-flavour balancing correlations as a complementary probe of heavy-flavour hadronisation. We compared the Monash, String Junctions (SJ) and Close-Packing (CP) configurations of PYTHIA-8 to investigate how charm and beauty quantum numbers are distributed among mesons and baryons produced in pp collisions. The integrated balancing yields show substantial differences between the configurations, particularly in the baryonic channels. These results demonstrate the sensitivity of heavy-flavour balancing observables to the hadronisation descriptions implemented in the three configurations. We found that the Monash tune exhibits a relatively weak multiplicity dependence. On the other hand, the SJ and CP tunes predict an increasing baryonic contribution in high-multiplicity events, consistent with the influence of junction formation and hadronisation in a dense string environment. These results highlight the potential interplay between multiple parton interactions, colour reconnection and heavy-flavour hadronisation. Future measurements by current but also future LHC experiments, benefiting from improved tracking, extended detector acceptance, particle identification and larger data samples, could test these model expectations through species-resolved heavy-flavour correlations as a function of event activity. Such measurements would provide new constraints on non-perturbative QCD models and help clarify the mechanisms responsible for meson and baryon production in high-energy pp collisions.

%% file: sections/Appendices.tex
\section{Additional hadronisation settings}
\label{sec:additional-hadronisation-settings}
\begin{table}[htbp]
    \centering
    \small
    \caption{
        PYTHIA settings used in the hadronisation configurations.
        CP-specific entries report the settings applied in that
        configuration; they do not imply different values in every other
        configuration. The common transverse-momentum width
        \texttt{StringPT:sigma} is the baseline value before local
        close-packing modifications.
    }
    \label{tab:additional-hadronisation-settings}

    \begin{tabular}{@{}lc@{}}
        \toprule
        \textbf{Parameter} & \textbf{Value} \\
        \midrule

        \multicolumn{2}{@{}l}{
            \textit{Additional CP settings}
        } \\[2pt]

        \texttt{ClosePacking:enhanceStrange}
            & 0.014 \\

        \texttt{ClosePacking:enhancePT}
            & 0.014 \\

        \texttt{ClosePacking:PT0}
            & $2.0\,\mathrm{GeV}$ \\

        \texttt{ClosePacking:doEnhanceDiquark}
            & \texttt{off} \\

        \texttt{ClosePacking:baryonSup}
            & 0.928 \\

        \texttt{ClosePacking:parallelBaryonSup}
            & 0 \\

        \texttt{StringFragmentation:doStrangeJunctions}
            & \texttt{on} \\

        \texttt{StringFragmentation:enhanceStrangeJunction}
            & 0.540 \\

        \texttt{Ropewalk:RopeHadronization}
            & \texttt{off} \\

        \midrule

        \multicolumn{2}{@{}l}{
            \textit{Shared by SJ and CP}
        } \\[2pt]

        \texttt{ColourReconnection:timeDilationMode}
            & 2 \\

        \texttt{ColourReconnection:timeDilationPar}
            & 0.18 \\

        \texttt{ColourReconnection:allowJunctions}
            & \texttt{on} \\

        \texttt{ColourReconnection:allowDoubleJunRem}
            & \texttt{off} \\

        \midrule

        \multicolumn{2}{@{}l}{
            \textit{Shared baseline for all three configurations}
        } \\[2pt]

        \texttt{StringPT:sigma}
            & $0.335\,\mathrm{GeV}$ \\

        \bottomrule
    \end{tabular}
\end{table}

\begin{table}[htbp]
    \centering
    \caption{Charged-particle multiplicity intervals corresponding to the
    event-activity percentile classes for the Monash, SJ, and
    CP configurations.}
    \label{tab:multiplicity_classes}

    \small
    \setlength{\tabcolsep}{8pt}
    \renewcommand{\arraystretch}{1.1}

    \begin{tabular}{cccc}
        \toprule
        \textbf{Percentile}
        & \textbf{Monash}
        & \textbf{SJ}
        & \textbf{CP} \\
        \midrule

        0--1     & $\geq 165$ & $\geq 173$ & $\geq 162$ \\
        1--10    & 98--164    & 105--172   & 98--161    \\
        10--20   & 71--97     & 77--104    & 72--97     \\
        20--30   & 54--70     & 60--76     & 57--71     \\
        30--40   & 44--53     & 49--59     & 46--56     \\
        40--50   & 35--43     & 40--48     & 38--45     \\
        50--60   & 28--34     & 33--39     & 30--37     \\
        60--70   & 22--27     & 26--32     & 24--29     \\
        70--80   & 18--21     & 20--25     & 19--23     \\
        80--90   & 14--17     & 15--19     & 14--18     \\
        90--100  & 0--13      & 0--14      & 0--13      \\

        \bottomrule
    \end{tabular}
\end{table}

\section{Statistical uncertainties and covariance}
\label{sec:stat-uncertainty}

In the following, we describe how we construct the statistical uncertainties using a delete-one-block jackknife. Each tune $u$ is divided into $K=10$ blocks, indexed by $b$, each containing ten million successful events from 100 independently seeded source jobs. Then, let $\mathbf X_{ub}$ collect the additive event, trigger and pair counts. The analysis function $g$ will map the pooled counts, kept separate for each tune, to the vector of central observable values $\hat{\boldsymbol\theta}$:

\begin{equation}
    \hat{\boldsymbol{\theta}}
    =
    g\left(
        \left\{
            \sum_{b=1}^{K}\mathbf{X}_{ub}
        \right\}_{u}
    \right).
    \label{eq:central-values}
\end{equation}

We proceed to remove block $b$ from tune $u$, keeping the other tunes complete, and then recompute the affected normalisations, ratios and
multiplicity-class boundaries. We can then denote the resulting observable vector by $\boldsymbol{\theta}_{u,-b}$, and its mean over the $K$ deletions will then be
\begin{equation}
    \bar{\boldsymbol{\theta}}_{u}
    =
    \frac{1}{K}
    \sum_{b=1}^{K}\boldsymbol{\theta}_{u,-b}.
    \label{eq:mean-leave-one-block-out}
\end{equation}
This deletion mean is used to estimate the covariance, while the reported central values remain those of
eq.~\eqref{eq:central-values}.

Because the tune samples are independent, their covariance contributions are added as follows:
\begin{equation}
    \widehat{\mathbf{C}}
    =
    \sum_{u}\frac{K-1}{K}
    \sum_{b=1}^{K}
    \left(
        \boldsymbol{\theta}_{u,-b}
        -
        \bar{\boldsymbol{\theta}}_{u}
    \right)
    \left(
        \boldsymbol{\theta}_{u,-b}
        -
        \bar{\boldsymbol{\theta}}_{u}
    \right)^{\mathrm{T}},
    \label{eq:jackknife-covariance}
\end{equation}
where $\mathrm{T}$ denotes the transpose and $(K-1)/K$ is the jackknife normalisation factor. Recomputing the observables together retains correlations between OS and SS counts, numerators and denominators, and quantities sharing a reference. Finally, the standard error of observable component $i$ is 
\begin{equation}
    \mathrm{SE}_{i}
    =
    \sqrt{\widehat{C}_{ii}},
    \label{eq:standard-error}
\end{equation}
where $\widehat{C}_{ii}$ is its diagonal covariance element.

Unresolved or non-finite denominators, invalid deletions, changing integer boundaries or unresolved threshold margins can prevent an SE from being reported. We thus distinguish an undefined central value from a finite value without an accepted SE.

Statistical errors exclude model/parameter systematics and do not automatically establish Gaussian confidence-interval coverage. Ten deletions also limit covariance precision and rank.

%% file: hfBalancingModelPaperRefs.bib
@book{Andersson:1997xwk,
    author = "Andersson, Bo",
    title = "{The Lund Model}",
    doi = "10.1017/9781009401296",
    isbn = "978-1-009-40129-6, 978-1-009-40125-8, 978-1-009-40128-9, 978-0-521-01734-3, 978-0-521-42094-5, 978-0-511-88149-7",
    publisher = "Cambridge University Press",
    volume = "7",
    year = "1998"
}

@article{ALICE:2025ytq,
    author = "Abualrob, Ibrahim Jaser and others",
    collaboration = "ALICE",
    title = "{Multiplicity dependence of two-particle angular correlations of identified particles in pp collisions at $\mathbf {\sqrt{s} = 13}$ TeV}",
    eprint = "2511.10399",
    archivePrefix = "arXiv",
    primaryClass = "nucl-ex",
    reportNumber = "CERN-EP-2025-253",
    doi = "10.1140/epjc/s10052-026-15447-z",
    journal = "Eur. Phys. J. C",
    volume = "86",
    number = "7",
    pages = "756",
    year = "2026"
}

@article{ALICE:2020wla,
    author = "Acharya, Shreyasi and others",
    collaboration = "ALICE",
    title = "{$\Lambda^+_c$ production in $pp$ and in $p$-Pb collisions at $\sqrt {s_{NN}}$=5.02 TeV}",
    eprint = "2011.06079",
    archivePrefix = "arXiv",
    primaryClass = "nucl-ex",
    reportNumber = "CERN-EP-2020-218",
    doi = "10.1103/PhysRevC.104.054905",
    journal = "Phys. Rev. C",
    volume = "104",
    number = "5",
    pages = "054905",
    year = "2021"
}

@article{ALICE:2021npz,
    author = "Acharya, Shreyasi and others",
    collaboration = "ALICE",
    title = "{Observation of a multiplicity dependence in the pT-differential charm baryon-to-meson ratios in proton{\textendash}proton collisions at s=13 TeV}",
    eprint = "2111.11948",
    archivePrefix = "arXiv",
    primaryClass = "nucl-ex",
    reportNumber = "CERN-EP-2021-245",
    doi = "10.1016/j.physletb.2022.137065",
    journal = "Phys. Lett. B",
    volume = "829",
    pages = "137065",
    year = "2022"
}

@article{Andersson:1983ia,
    author = "Andersson, Bo and Gustafson, G. and Ingelman, G. and Sjostrand, T.",
    title = "{Parton Fragmentation and String Dynamics}",
    reportNumber = "LU-TP-83-10",
    doi = "10.1016/0370-1573(83)90080-7",
    journal = "Phys. Rept.",
    volume = "97",
    pages = "31--145",
    year = "1983"
}

@inproceedings{Andersson:2002ap,
    author = "Andersson, Bo and Mohanty, Sandipan and Soderberg, Fredrik",
    title = "{Recent developments in the Lund model}",
    booktitle = "{36th Annual Winter School on Nuclear and Particle Physics (PINP 2002) and 8th St. Petersburg School on Theoretical Physics}",
    eprint = "hep-ph/0212122",
    archivePrefix = "arXiv",
    reportNumber = "LU-TP-02-45",
    month = "12",
    year = "2002"
}

@article{Christiansen:2015yqa,
    author = "Christiansen, Jesper R. and Skands, Peter Z.",
    title = "{String Formation Beyond Leading Colour}",
    eprint = "1505.01681",
    archivePrefix = "arXiv",
    primaryClass = "hep-ph",
    reportNumber = "COEPP-MN-15-1, LU-TP-15-16, MCNET-15-09",
    doi = "10.1007/JHEP08(2015)003",
    journal = "JHEP",
    volume = "08",
    pages = "003",
    year = "2015"
}

@article{Collins:1989gx,
    author = "Collins, John C. and Soper, Davison E. and Sterman, George F.",
    title = "{Factorization of Hard Processes in QCD}",
    eprint = "hep-ph/0409313",
    archivePrefix = "arXiv",
    reportNumber = "ITP-SB-89-31",
    doi = "10.1142/9789814503266_0001",
    journal = "Adv. Ser. Direct. High Energy Phys.",
    volume = "5",
    pages = "1--91",
    year = "1989"
}

@article{Metz:2016swz,
    author = "Metz, Andreas and Vossen, Anselm",
    title = "{Parton Fragmentation Functions}",
    eprint = "1607.02521",
    archivePrefix = "arXiv",
    primaryClass = "hep-ex",
    doi = "10.1016/j.ppnp.2016.08.003",
    journal = "Prog. Part. Nucl. Phys.",
    volume = "91",
    pages = "136--202",
    year = "2016"
}

@inproceedings{Webber:1994zd,
    author = "Webber, B. R.",
    title = "{Hadronization}",
    booktitle = "{Summer School on Hadronic Aspects of Collider Physics}",
    eprint = "hep-ph/9411384",
    archivePrefix = "arXiv",
    reportNumber = "CAVENDISH-HEP-94-17",
    pages = "49--77",
    month = "11",
    year = "1994"
}

@article{Sjostrand:2014zea,
    author = {Sj{\"o}strand, Torbj{\"o}rn and Ask, Stefan and Christiansen, Jesper R. and Corke, Richard and Desai, Nishita and Ilten, Philip and Mrenna, Stephen and Prestel, Stefan and Rasmussen, Christine O. and Skands, Peter Z.},
    title = "{An introduction to PYTHIA 8.2}",
    eprint = "1410.3012",
    archivePrefix = "arXiv",
    primaryClass = "hep-ph",
    reportNumber = "LU-TP-14-36, MCNET-14-22, CERN-PH-TH-2014-190, FERMILAB-PUB-14-316-CD, DESY-14-178, SLAC-PUB-16122",
    doi = "10.1016/j.cpc.2015.01.024",
    journal = "Comput. Phys. Commun.",
    volume = "191",
    pages = "159--177",
    year = "2015"
}

@article{Bierlich:2015rha,
    author = "Bierlich, Christian and Christiansen, Jesper Roy",
    title = "{Effects of color reconnection on hadron flavor observables}",
    eprint = "1507.02091",
    archivePrefix = "arXiv",
    primaryClass = "hep-ph",
    reportNumber = "LU-TP-15-26, MCNET-15-16",
    doi = "10.1103/PhysRevD.92.094010",
    journal = "Phys. Rev. D",
    volume = "92",
    number = "9",
    pages = "094010",
    year = "2015"
}

@article{ALICE:2016fzo,
    author = "Adam, Jaroslav and others",
    collaboration = "ALICE",
    title = "{Enhanced production of multi-strange hadrons in high-multiplicity proton-proton collisions}",
    eprint = "1606.07424",
    archivePrefix = "arXiv",
    primaryClass = "nucl-ex",
    reportNumber = "CERN-EP-2016-153",
    doi = "10.1038/nphys4111",
    journal = "Nature Phys.",
    volume = "13",
    pages = "535--539",
    year = "2017"
}

@article{Altmann:2025afh,
    author = "Altmann, Javira and Bernardinis, Lorenzo and Skands, Peter and Zaccolo, Valentina",
    title = "{Closepacking effects on strangeness and baryon production at the LHC}",
    eprint = "2512.00671",
    archivePrefix = "arXiv",
    primaryClass = "hep-ph",
    doi = "10.1007/JHEP04(2026)164",
    journal = "JHEP",
    volume = "04",
    pages = "164",
    year = "2026"
}

@article{Skands:2014pea,
  author = {Skands, Peter and Carrazza, Stefano and Rojo, Juan},
  title = {{Tuning PYTHIA 8.1: the Monash 2013 Tune}},
  eprint = {1404.5630},
  archivePrefix = {arXiv},
  primaryClass = {hep-ph},
  doi = {10.1140/epjc/s10052-014-3024-y},
  journal = {Eur. Phys. J. C},
  volume = {74},
  number = {8},
  pages = {3024},
  year = {2014},
  url = {https://doi.org/10.1140/epjc/s10052-014-3024-y},
}

@article{ALICE:2021rzj,
  author = {Acharya, Shreyasi and others},
  collaboration = {ALICE},
  title = {{Measurement of Prompt $D^0$, $\Lambda_c^+$, and $\Sigma_c^{0,++}(2455)$ Production in Proton--Proton Collisions at $\sqrt{s}=13\,\mathrm{TeV}$}},
  eprint = {2106.08278},
  archivePrefix = {arXiv},
  primaryClass = {hep-ex},
  doi = {10.1103/PhysRevLett.128.012001},
  journal = {Phys. Rev. Lett.},
  volume = {128},
  number = {1},
  pages = {012001},
  year = {2022},
  url = {https://doi.org/10.1103/PhysRevLett.128.012001},
}

@article{Bierlich:2022pfr,
  author = {Bierlich, Christian and others},
  title = {{A comprehensive guide to the physics and usage of PYTHIA 8.3}},
  eprint = {2203.11601},
  archivePrefix = {arXiv},
  primaryClass = {hep-ph},
  doi = {10.21468/SciPostPhysCodeb.8},
  journal = {SciPost Phys. Codeb.},
  volume = {2022},
  pages = {8},
  year = {2022},
  url = {https://doi.org/10.21468/SciPostPhysCodeb.8},
}

@article{LHCb:2023wbo,
  author = {Aaij, Roel and others},
  collaboration = {LHCb},
  title = {{Enhanced Production of $\Lambda_b^0$ Baryons in High-Multiplicity $pp$ Collisions at $\sqrt{s}=13\,\mathrm{TeV}$}},
  eprint = {2310.12278},
  archivePrefix = {arXiv},
  primaryClass = {hep-ex},
  doi = {10.1103/PhysRevLett.132.081901},
  journal = {Phys. Rev. Lett.},
  volume = {132},
  number = {8},
  pages = {081901},
  year = {2024},
  url = {https://doi.org/10.1103/PhysRevLett.132.081901},
}

@article{Pruneau:2024jpa,
  author = {Pruneau, Claude and Basu, Sumit and Gonzalez, Victor and Hanley, Brian and Marin, Ana and Dobrin, Alexandru F. and Manea, Alexandru},
  title = {{Mixed-species charge and baryon balance functions studies with PYTHIA}},
  eprint = {2403.13007},
  archivePrefix = {arXiv},
  primaryClass = {hep-ph},
  doi = {10.1103/PhysRevC.109.064913},
  journal = {Phys. Rev. C},
  volume = {109},
  number = {6},
  pages = {064913},
  year = {2024},
  url = {https://doi.org/10.1103/PhysRevC.109.064913},
}

@article{Sheibani:2025iuh,
  author = {Sheibani, Oveis and Pruneau, Claude and Gonzalez, Victor and Basu, Sumit and Dobrin, Alexandru Florin and Patley, Yash and Nandi, Basanta and Dash, Sadhana},
  title = {{Probing partonic evolution and hadronization via balance functions and correlations of charmed hadrons}},
  eprint = {2507.21014},
  archivePrefix = {arXiv},
  primaryClass = {hep-ex},
  doi = {10.1103/7j49-x6xf},
  journal = {Phys. Rev. C},
  volume = {112},
  number = {6},
  pages = {065210},
  year = {2025},
  url = {https://doi.org/10.1103/7j49-x6xf},
}

@article{Bass:2000az,
  author = {Bass, Steffen A. and Danielewicz, Pawel and Pratt, Scott},
  title = {{Clocking hadronization in relativistic heavy ion collisions with balance functions}},
  eprint = {nucl-th/0005044},
  archivePrefix = {arXiv},
  doi = {10.1103/PhysRevLett.85.2689},
  journal = {Phys. Rev. Lett.},
  volume = {85},
  pages = {2689--2692},
  year = {2000},
  url = {https://doi.org/10.1103/PhysRevLett.85.2689},
}

@article{
Gross:2005kv,
author = {David J. Gross },
title = {{The discovery of asymptotic freedom and the emergence of QCD}},
journal = {Proceedings of the National Academy of Sciences},
volume = {102},
number = {26},
pages = {9099-9108},
year = {2005},
doi = {10.1073/pnas.0503831102},
URL = {https://www.pnas.org/doi/abs/10.1073/pnas.0503831102}}

@article{Politzer:1973fx,
  title = {{Reliable Perturbative Results for Strong Interactions?}},
  author = {Politzer, H. David},
  journal = {Phys. Rev. Lett.},
  volume = {30},
  issue = {26},
  pages = {1346--1349},
  numpages = {0},
  year = {1973},
  month = {Jun},
  publisher = {American Physical Society},
  doi = {10.1103/PhysRevLett.30.1346},
  url = {https://link.aps.org/doi/10.1103/PhysRevLett.30.1346}
}

@book{Ioffe:2010zz, place={Cambridge}, series={Cambridge Monographs on Particle Physics, Nuclear Physics and Cosmology}, title={{Quantum Chromodynamics: Perturbative and Nonperturbative Aspects}}, publisher={Cambridge University Press}, author={Ioffe, B. L. and Fadin, V. S. and Lipatov, L. N.}, year={2010}, collection={Cambridge Monographs on Particle Physics, Nuclear Physics and Cosmology}}

@book{Greensite:2011zz,
    author    = {Jeff Greensite},
    title     = {{An Introduction to the Confinement Problem}},
    publisher = {Springer},
    series    = {Lecture Notes in Physics},
    volume    = {821},
    year      = {2011},
    address   = {Berlin, Heidelberg},
    doi       = {10.1007/978-3-642-14382-3},
    isbn      = {978-3-642-14382-3}
}

@article{ALICE:2023wbx,
    title         = {{Study of Flavor Dependence of the Baryon-to-Meson Ratio in Proton-Proton Collisions at $\sqrt{s}=13$ TeV}},
    author        = {{ALICE Collaboration}},
    journal       = {Physical Review D},
    volume        = {108},
    number        = {11},
    pages         = {112003},
    year          = {2023},
    month         = dec,
    doi           = {10.1103/PhysRevD.108.112003},
    url           = {https://doi.org/10.1103/PhysRevD.108.112003}
}

@article{ALICE:2020wfu,
    title         = {{$\Lambda_c^+$ Production and Baryon-to-Meson Ratios in $pp$ and $p$--Pb Collisions at $\sqrt{s_{\mathrm{NN}}}=5.02$ TeV at the LHC}},
    author        = {{ALICE Collaboration}},
    journal       = {Physical Review Letters},
    volume        = {127},
    number        = {20},
    pages         = {202301},
    year          = {2021},
    month         = nov,
    doi           = {10.1103/PhysRevLett.127.202301},
    url           = {https://doi.org/10.1103/PhysRevLett.127.202301}
}

@article{ALICE:2017thy,
    title         = {{$\Lambda_c^+$ Production in $pp$ Collisions at $\sqrt{s}=7$ TeV and in $p$--Pb Collisions at $\sqrt{s_{\mathrm{NN}}}=5.02$ TeV}},
    author        = {{ALICE Collaboration}},
    journal       = {Journal of High Energy Physics},
    volume        = {2018},
    number        = {4},
    pages         = {108},
    year          = {2018},
    month         = apr,
    doi           = {10.1007/JHEP04(2018)108},
    url           = {https://doi.org/10.1007/JHEP04(2018)108}
}

@article{ALICE:2021psx,
    title         = {{Measurement of the Production Cross Section of Prompt $\Xi_c^0$ Baryons at Midrapidity in $pp$ Collisions at $\sqrt{s}=5.02$ TeV}},
    author        = {{ALICE Collaboration}},
    journal       = {Journal of High Energy Physics},
    volume        = {2021},
    number        = {10},
    pages         = {159},
    year          = {2021},
    month         = oct,
    doi           = {10.1007/JHEP10(2021)159},
    url           = {https://doi.org/10.1007/JHEP10(2021)159}
}

@article{ALICE:2021bli,
    title         = {{Measurement of the Cross Sections of $\Xi_c^0$ and $\Xi_c^+$ Baryons and of the Branching-Fraction Ratio $\mathrm{BR}(\Xi_c^0\to\Xi^-e^+\nu_e)/\mathrm{BR}(\Xi_c^0\to\Xi^-\pi^+)$ in $pp$ Collisions at $\sqrt{s}=13$ TeV}},
    author        = {{ALICE Collaboration}},
    journal       = {Physical Review Letters},
    volume        = {127},
    number        = {27},
    pages         = {272001},
    year          = {2021},
    month         = dec,
    doi           = {10.1103/PhysRevLett.127.272001},
    url           = {https://doi.org/10.1103/PhysRevLett.127.272001}
}

@article{ALICE:2017dja,
    title         = {{First Measurement of $\Xi_c^0$ Production in $pp$ Collisions at $\sqrt{s}=7$ TeV}},
    author        = {{ALICE Collaboration}},
    journal       = {Physics Letters B},
    volume        = {781},
    pages         = {8--19},
    year          = {2018},
    month         = jun,
    doi           = {10.1016/j.physletb.2018.03.061},
    url           = {https://doi.org/10.1016/j.physletb.2018.03.061}
}

@article{ALICE:2022cop,
    title         = {{First Measurement of $\Omega_c^0$ Production in $pp$ Collisions at $\sqrt{s}=13$ TeV}},
    author        = {{ALICE Collaboration}},
    journal       = {Physics Letters B},
    volume        = {846},
    pages         = {137625},
    year          = {2023},
    month         = sep,
    doi           = {10.1016/j.physletb.2022.137625},
    url           = {https://doi.org/10.1016/j.physletb.2022.137625}
}

@article{CMS:2016zzh,
    title         = {{Multiplicity and Rapidity Dependence of Strange Hadron Production in $pp$, $p$Pb, and PbPb Collisions at the LHC}},
    author        = {Khachatryan, V. and others},
    journal       = {Physics Letters B},
    volume        = {768},
    pages         = {103--129},
    year          = {2017},
    month         = may,
    doi           = {10.1016/j.physletb.2017.01.075},
    url           = {https://doi.org/10.1016/j.physletb.2017.01.075}
}

@article{ATLAS:2024nbm,
    author        = {{ATLAS Collaboration}},
    title         = {{Underlying-Event Studies with Strange Hadrons in $pp$ Collisions at $\sqrt{s}=13$ TeV with the ATLAS Detector}},
    journal       = {The European Physical Journal C},
    volume        = {84},
    pages         = {1335},
    year          = {2024},
    doi           = {10.1140/epjc/s10052-024-13243-1},
    url           = {https://doi.org/10.1140/epjc/s10052-024-13243-1}
}

@article{Lonnblad:2023kft,
    author        = {L{\"o}nnblad, Leif and Shah, Harsh},
    title         = {{Baryon Correlations in Pythia}},
    eprint        = {2309.01557},
    archivePrefix = {arXiv},
    primaryClass  = {hep-ph},
    reportNumber  = {LU-TP-23-12, MCNET-23-08, MCNET-23-07},
    doi           = {10.1140/epjc/s10052-023-12271-7},
    journal       = {European Physical Journal C},
    volume        = {83},
    number        = {12},
    pages         = {1105},
    year          = {2023}
}

@article{ALICE:2020msa,
    author = "Acharya, Shreyasi and others",
    collaboration = "ALICE",
    title = "{Multiplicity dependence of J/$\psi$ production at midrapidity in pp collisions at $\sqrt{s}$ = 13 TeV}",
    eprint = "2005.11123",
    archivePrefix = "arXiv",
    primaryClass = "nucl-ex",
    reportNumber = "CERN-EP-2020-088",
    doi = "10.1016/j.physletb.2020.135758",
    journal = "Phys. Lett. B",
    volume = "810",
    pages = "135758",
    year = "2020"
}
